# Horizontal line nodes in Sr$_2$RuO$_4$ revealed by spin resonance

Kazuki Iida[1, 2], Maiko Kofu[1, 3], Katsuhiro Suzuki[4], Naoki Murai[3], Seiko Ohira-Kawamura[3], Ryoichi Kajimoto[3], Yasuhiro Inamura[3], Motoyuki Ishikado[2], Shunsuke Hasegawa[5], Takatsugu Masuda[5], Yoshiyuki Yoshida[6], Kazuhisa Kakurai[2], Kazushige Machida[7], & Seunghun Lee[1]

[1]Department of Physics, University of Virginia, Charlottesville, Virginia 22904-4714, USA
[2]Neutron Science and Technology Center, Comprehensive Research Organization for Science and Society (CROSS), Tokai, Ibaraki 319-1106, Japan
[3]J-PARC Center, Japan Atomic Energy Agency (JAEA), Tokai, Ibaraki 319-1195, Japan
[4]Research Organization of Science and Technology, Ritsumeikan University, Kusatsu, Shiga 525-8577, Japan
[5]Neutron Science Laboratory, Institute for Solid State Physics, University of Tokyo, Kashiwa, Chiba 277-8581, Japan
[6]National Institute of Advanced Industrial Science and Technology, Tsukuba, Ibaraki 305-8565, Japan
[7]Department of Physics, Ritsumeikan University, Kusatsu, Shiga 525-8577, Japan

Low-energy incommensurate (IC) magnetic fluctuations in Sr$_2$RuO$_4$ are investigated by the high-resolution inelastic neutron scattering measurements and random phase approximation (RPA) calculations. Analysis of the $L$ dependences of the low-energy IC magnetic fluctuations enables us to observe the spin resonance at $\mathbf{Q}_{res}$ = (0.3, 0.3, 0.5) and $\hbar\omega_{res}$ = 0.56 meV which corresponds well to the superconducting gap $2\Delta$ = 0.56 meV estimated by the tunneling spectroscopy. The spin resonance shows the $L$ modulated intensity with maximum at around $L$ = 0.5. The $L$ modulated intensity of the spin resonance and our RPA calculations demonstrate that the superconducting gaps regarding the quasi-one-dimensional $\alpha$ and $\beta$ sheets at the Fermi surfaces have the horizontal line nodes. The present results can give a strong constraint on the pairing symmetry of Sr$_2$RuO$_4$, and possible superconducting order parameters are discussed.

**Introduction**

Strontium ruthenate $Sr_2RuO_4$ with $T_c$ = 1.5 K[1] has attracted a great deal of interest as the prime candidate for a chiral *p*-wave superconductor[2-5]; nuclear magnetic resonance (NMR)[6] and polarized neutron scattering[7] measurements reported spin-triplet superconductivity whereas muon spin rotation[8] and Kerr effect[9] measurements showed spontaneously time reversal symmetry breaking. On the other hand, there are some experimental results such as absence of the chiral edge currents[10], first-order superconducting transition[11-14], and strong $H_{c2}$ (∥ *ab*) suppression[15], all of which challenge the chiral *p*-wave superconductivity. Recently, experimental and theoretical studies under an application of uniaxial pressure along ⟨100⟩ reported a factor of 2.3 enhancement of $T_c$ owing to the Lifshitz transition when the Fermi level passes through a van Hove singularity, raising the possibility of an even-parity spin-singlet order parameter in $Sr_2RuO_4$[16-22]. More recently, new NMR results demonstrated that the spin susceptibility substantially drops below $T_c$ provided that the pulse energy is smaller than a threshfold[23,24], strongly contradicting the chiral-*p* spin-triplet superconductivity. As such, these recent works turn the research on $Sr_2RuO_4$ towards a fascinating new phase.

So far, various experimental techniques reported that the superconducting gaps of $Sr_2RuO_4$ have line nodes[25,26], but the details of the line nodes, e.g. vertical or horizontal, are not uncovered yet. The thermal conductivity measurements reported vertical line nodes on the superconducting gaps[27], whereas field-angle-dependent specific heat capacity measurements reported the horizontal line nodes[28,29]. Since the complete information of the superconducting gaps can shed light on the pairing symmetry, exclusive determination of the direction of the line nodes in $Sr_2RuO_4$ is now called for.

Inelastic neutron scattering (INS) technique can directly measure imaginary part of generalized spin susceptibility ($\chi''$) as a function of momentum (**Q**) and energy ($\hbar\omega$) transfers, yielding abundant information of Fermi surface topology. In addition, **Q** dependence of the spin resonance as a consequence of the Bardeen-Cooper-Schrieffer (BCS) coherence factor can provide the information on the symmetry of superconducting gaps. In $Sr_2RuO_4$, the most pronounced magnetic signal in the normal state is nearly two-dimensional incommensurate (IC) magnetic fluctuations at $\mathbf{Q}_{IC}$ = (0.3, 0.3, *L*) owing to the Fermi surface nesting between (or within) the quasi-one-dimensional *α* and *β* sheets consisted of the $d_{zx}$ and $d_{yz}$ orbitals of $Ru^{4+}$ [30-33]. In contrast to the pronounced signal from the IC magnetic fluctuations, only blurred signals due to ferromagnetic fluctuations originating from the two-dimensional *γ* sheet ($d_{xy}$) are observed around the Γ point[34,35]. Upon decreasing temperature below $T_c$, no sizable spin resonance regarding the IC magnetic fluctuations[36] was, however, observed at **Q** = (0.3, 0.3, 0) and the energy close to the superconducting gap 2Δ = 0.56 meV[33] estimated by the tunneling spectroscopy[37].

Based on the horizontal line nodes model, the spin resonance is supposed to emerge at **Q** = (0.3, 0.3, *L*) with finite *L* but not at (0.3, 0.3, 0) because the sign of the superconducting gap changes along $k_z$[29]. In the meanwhile, the vertical line nodes model expects the spin resonance at **Q** = (0.3, 0.3, 0) since the sign changes in the $k_z$ plane. Therefore, the horizontal line nodes model can naturally explain the absence of the spin resonance at **Q** = (0.3, 0.3,

0), and search for the spin resonance at (0.3, 0.3, L) is of particular interest. In this paper, we investigate in detail the low-energy IC magnetic fluctuations in Sr$_2$RuO$_4$ by analyzing the L dependences using high-resolution INS technique. The experimental results, especially the L dependence of neutron scattering intensities, are compared with the random phase approximation (RPA) calculations.

## Results
### High-energy INS results.
Overall features of high-energy ($\hbar\omega \gg 2\Delta$) IC magnetic fluctuations with $L = 0.5$ in Sr$_2$RuO$_4$ are summarized in Fig. 1. Figure 1a depicts the constant-energy INS intensity map in the ($HK$0.5) plane at 0.3 K and 2.5 meV. IC magnetic fluctuations are observed at $\mathbf{Q}_{IC}$ = (0.3, 0.3, 0.5), (0.7, 0.3, 0.5), and (0.7, 0.7, 0.5). In addition to the IC magnetic peaks, the Fermi surface nesting also induces the ridge scattering connecting $\mathbf{Q}_{IC}$ around (0.5, 0.5, 0.5). To explore the energy evolution of the IC magnetic fluctuations with $L = 0.5$, the INS intensity is converted to the imaginary part of the spin susceptibility $\chi''(\hbar\omega)$ via the fluctuation dissipation theorem $\chi''(\hbar\omega) = (1 - e^{-\hbar\omega/k_B T})I(\hbar\omega)$ after subtracting the background. The $\chi''(\hbar\omega)$ spectra at $\mathbf{Q}_{IC}$ = (0.3, 0.3, 0.5) below and above $T_c$ are plotted in Fig. 1c. The $\chi''(\hbar\omega)$ spectra above 1.0 meV are well fitted by the relaxation response model $\chi''(\hbar\omega) = \chi'\Gamma\hbar\omega/[(\hbar\omega)^2 + \Gamma^2]$ where $\chi'$ is the static susceptibility and $\Gamma$ the relaxation rate [or the peak position of $\chi''(\hbar\omega)$], yielding $\Gamma$ = 6.3(2) meV [6.5(2) meV] at 0.3 K (1.8 K). Therefore, the observed IC magnetic fluctuations with $L = 0.5$ are quantitatively consistent with the IC magnetic fluctuations with $L = 0$ reported in the previous INS works[31-34]. This is reasonable since the IC magnetic fluctuations along (0.3, 0.3, L) monotonically decreases in intensity with increasing L (Fig. 1b), representing the quasi-two-dimensional feature of the IC magnetic fluctuations in this energy region[34] consistent with the quasi-one-dimensional band structures of the cylindrical α and β sheets[38].

### Low-energy INS results.
In the following, we concentrate on the low-energy ($\hbar\omega \simeq 2\Delta$) IC magnetic fluctuations of Sr$_2$RuO$_4$. As described below, in the superconducting state, a spin resonance appears at $\mathbf{Q}_{res}$ = (0.3, 0.3, 0.5) and $\hbar\omega_{res} = 2\Delta$ (white arrow in Fig. 2a) but not at $\mathbf{Q}$ = (0.3, 0.3, 0) (Fig. 2c), as expected for the horizontal line nodes model. The L modulated intensity of the spin resonance is only observed at $\hbar\omega = 2\Delta$ (Fig. 2e). The intensity modulation along L is in sharp contrast with the nearly two-dimensional IC magnetic fluctuations at high energy transfer (Fig. 1b), indicating the presence of three-dimensional superconducting gaps. For quantitative analysis on the spin resonance, $\hbar\omega$ and $\mathbf{Q}$ dependences of INS intensities [$I(\hbar\omega)$ and $I(\mathbf{Q})$] are investigated.

$I(\hbar\omega)$ cuts at $\mathbf{Q}$ = (0.3, 0.3, 0.5) and (0.3, 0.3, 0) are plotted in Fig. 3a, b, respectively. Below $T_c$, a clear increase in intensity at $\hbar\omega \sim 0.56$ meV can be seen in the $I(\hbar\omega)$ cut at (0.3, 0.3, 0.5) (solid area in Fig. 3a). Meanwhile, such enhancement at $\hbar\omega \sim 0.56$ meV is not observed in the $I(\hbar\omega)$ cut at (0.3, 0.3, 0) (Fig. 3b) as reported in the previous INS study[33]. $I(\mathbf{Q})$ cuts at $\hbar\omega = 0.56$ meV along $\mathbf{Q}$ = ($H, H, 0.5$) and ($H, H, 0$) also show the same trend (Fig. 3c, d). The IC peak at $\hbar\omega = 0.56$ meV along $\mathbf{Q}$ = ($H, H, 0.5$) is enhanced below $T_c$ (solid area in Fig.

3c), while the IC peak at $\hbar\omega$ = 0.56 meV along **Q** = (*H*, *H*, 0) do not change in intensity across $T_c$ (Fig. 3d). Energy evolution of the *I*(**Q**) cuts along **Q** = (*H*, *H*, 0.5) is plotted in Fig. 3e. Although *I*(**Q**) cuts show the peaks at $\mathbf{Q}_{IC}$ = (0.3, 0.3, 0.5) in all the energy windows at both temperatures, the INS intensity at $\hbar\omega$ = 0.56 meV only shows sizable enhancement at $\mathbf{Q}_{IC}$ in the superconducting state compared to the normal state (solid area in Fig. 3e). It should be emphasized that the energy corresponds well to the superconducting gap $2\Delta$[37] (vertical bar in Fig. 3a), indicating that the observed enhancement, localized in both $\mathbf{Q}_{res}$ = (0.3, 0.3, 0.5) and $\hbar\omega_{res}$ = 0.56 meV, is the spin resonance (see also white arrows in Fig. 2a-c). The spin gap of the IC magnetic fluctuations at $\mathbf{Q}_{IC}$ = (0.3, 0.3, 0.5) is smaller than 0.2 meV (Fig. 3a), which makes the spin resonance less pronounced.

*L* dependence of the low-energy IC magnetic fluctuations provides us the compelling evidence to clarify the line nodes at the superconducting gaps[29]. Contour maps of the INS intensities in the (*HHL*) plane at 0.3 K with the energies 0.35, 0.56, and 0.74 meV are illustrated in Fig. 2d-f, and *I*(**Q**) cuts along **Q** = (0.3, 0.3, *L*) at 0.3 K with the corresponding energies are also plotted in Fig. 3f. In contrast to the monotonically decreasing intensities of the IC magnetic fluctuations at 0.35 and 0.74 meV following the squared magnetic form factor of $Sr_2RuO_4$[39] (the dashed lines in Fig. 3f), the *I*(**Q**) cut at the spin resonance energy (0.56 meV) shows the maximum intensity at *L* ~ 0.5 besides the monotonically decreasing component of the IC magnetic fluctuations owing to the magnetic form factor (the solid area and the dashed line in Fig. 3f). The *L* modulated intensity is therefore the characteristic property of the spin resonance representing the gap symmetry as the feedback effect from the superconducting gaps. In addition, the *L* modulated intensity is the reason for the absence of the spin resonance at **Q** = (0.3, 0.3, 0) (Fig. 2c, 3b, 3d).

**RPA results.**
To theoretically elucidate the origin of the *L* modulated intensity of the spin resonance in $Sr_2RuO_4$, RPA calculations assuming the horizontal line nodes at the superconducting gaps [Eq. (1)] are performed. Calculated dynamical spin susceptibilities at energies $2\Delta_0$ and $4\Delta_0$ are shown in Fig. 4a, b. It should be mentioned that the squared magnetic form factor of $Sr_2RuO_4$[39] is not included in the current calculations. At $2\Delta_0$, the dynamical spin susceptibility shows the maximum at **Q** = (1/3, 1/3, 1/2) and (2/3, 2/3, 1/2) as illustrated in Fig. 4a. The superconducting gaps with the horizontal line nodes described in Eq. (1) give such the feature along *L*, which is indeed observed in our INS measurements (Fig. 2e). We address that the observed *L* modulated intensity cannot be explained by the superconducting gaps with the vertical line nodes[29]. On the other hand, any pronounced *L* modulation at higher energy transfers are observed neither in the experiment (Fig. 2f), except the monotonically decreasing intensities due to the magnetic form factor, nor in the calculation (Fig. 4b). Since the energy window is too high compared to the superconducting gaps in $Sr_2RuO_4$[37], the effect of the superconducting gaps is smeared out and only the quasi-two-dimensional feature of the IC magnetic fluctuations can be seen in the higher energy windows ($\hbar\omega \gg 2\Delta$).

**Discussion**
Because of the *L* analysis of the low-energy IC magnetic fluctuations below and above $T_c$,

we succeeded in observing the spin resonance at $\mathbf{Q}_{res}$ = (0.3, 0.3, 0.5) and $\hbar\omega_{res}$ = 2Δ in $Sr_2RuO_4$. This result indicates that the quasi-one-dimensional α and β sheets are active bands for the bulk superconductivity. Our RPA calculations reveal that the observed L modulated intensity of the spin resonance originates from the horizonal line nodes at the superconducting gaps, in agreement with the field-angle-dependent specific heat capacity measurements[28] and thermal conductivity measurements[40].

We now discuss the pairing symmetry of $Sr_2RuO_4$. Among the pairing symmetry with the horizontal line nodes proposed by the resent uniaxial pressure and NMR works[16-24], the chiral d-wave $(k_x + ik_y)k_z$[41] is compatible with the time reversal symmetry breaking[8,9]. However, its wave function is odd in the $k_z$ plane giving rise to the spin resonance at $\mathbf{Q}$ = (0.3, 0.3, 0), and thus such pairing symmetry can be excluded by the current INS measurements. Recently, time-reversal invariant superconductivity in $Sr_2RuO_4$ is proposed by the Josephson effects[42]. Combined with the time-reversal invariant superconductivity and the horizontal line nodes, d-wave $d_{3k_z^2-1}$ [29] or nodal s' is proposed for the candidate for the pairing symmetry of $Sr_2RuO_4$.

The single component order parameter is also in line with the absence of a split transition in the presence of uniaxial strain and $T_c$ cusp in the limit of zero strain[16-18]. Therefore, although INS technique is an insensitive probe for time reversal symmetry of superconductivity, the present results give a strong constraint on the superconducting pairing symmetry of $Sr_2RuO_4$.

In summary, we investigated the low-energy IC magnetic fluctuations in $Sr_2RuO_4$ below and above $T_c$ by analyzing the L dependence. Below $T_c$, the spin resonance appears at $\mathbf{Q}_{res}$ = (0.3, 0.3, 0.5) and $\hbar\omega_{res}$ = 0.56 meV. The spin resonance shows the L modulated intensity with maximum at L ~ 0.5. The L modulated intensity of the spin resonance and our RPA calculations indicate that the superconducting gaps regarding the α and β sheets have horizontal line nodes.

## Methods
### Synthesis of single crystals.
Three single crystals of $Sr_2RuO_4$ with a total mass of ~10 g were prepared by the floating-zone method[43,44], and each crystal shows $T_c$ ~ 1.4 K (onset). They were co-aligned in a way that the (HHL) plane is perpendicular to the rotating axis.

### INS measurements.
Time-of-flight neutron scattering measurements at 0.3 and 1.8 K were performed using the disk chopper spectrometer AMATERAS installed at Japan Proton Accelerator Research Complex[45,46]. The disk chopper was rotated at a frequency of 300 Hz, yielding the combinations of incident neutron energies 2.64, 5.93, and 23.7 meV with the energy resolutions 0.046, 0.146, and 1.07 meV, respectively, at the elastic channel.

### RPA calculations.
To construct a realistic model, we perform density functional theory (DFT) calculations using the Wien2k package[47]. We obtain an effective 3-orbital model considering the Ru $d_{xz}$, $d_{yz}$, $d_{xy}$-orbitals using the maximum localized Wannier functions[48]. The generalized gradient approximation (GGA) exchange-correlation functional[49] is adopted with the cut-off energy $RK_{max} = 7$ and 512 k-point mesh. We renormalize the bandwidth considering the effective mass $m^* = 3.5$, and the resulting renormalized bandwidth is $W \sim 1.05$ eV. We consider the following gap function with the horizontal line nodes:

$$\Delta(\mathbf{k}) = \Delta_0 \cos ck_z \tag{1}$$

within the standard BCS framework. We take the gap amplitude $\Delta_0 = 4.8 \times 10^{-3} W$. In the body center tetragonal system, the period along to the $k_z$ axis is $4\pi/c$, and thus we take $k_z$ as $0 < k_z < 4\pi/c$. We obtain the dynamical spin susceptibility $\chi_s^{total}(\mathbf{q}, \omega)$ applying RPA as

$$\chi_s^{total}(\mathbf{q}, \omega) = \sum_{l,m} \chi_s^{l,l,m,m}(\mathbf{q}, \omega) \tag{2}$$

$$\hat{\chi}_s(\mathbf{q}, \omega) = \hat{\chi}_0(\mathbf{q}, \omega)[\hat{I} - \hat{S}\hat{\chi}_0(\mathbf{q}, \omega)]^{-1} \tag{3}$$

$$\hat{\chi}_0(\mathbf{q}, \omega) = \hat{\chi}_{0,G}(\mathbf{q}, \omega) + \hat{\chi}_{0,F}(\mathbf{q}, \omega) \tag{4}$$

$$\chi_{0,G(F)}^{l_1,l_2,l_3,l_4}{}_{\sigma_1,\sigma_2,\sigma_3,\sigma_4}(\mathbf{q}, \omega) = \sum_k \sum_{n,m} \frac{f(E_{\mathbf{k+q}}^n) - f(E_{\mathbf{k}}^m)}{\omega + i\delta - E_{\mathbf{k+q}}^n + E_{\mathbf{k}}^n} \tag{5}$$

$$\times U_{l_1,\sigma_1,n}(\mathbf{k+q}) U_{l_4,\sigma_4,m}(\mathbf{k})$$

$$\times U^\dagger_{m,l_2,\sigma_2}(\mathbf{k}) U^\dagger_{n,l_3,\sigma_3}(\mathbf{k+q})$$

where $l_1 \sim l_4$ and $\sigma_1 \sim \sigma_4$ are the orbital ($d_{zx}$, $d_{yz}$, $d_{xy}$) and spin (↑ and ↓) indices. $\hat{\chi}_{0,G(F)}$ denotes the normal (anomalous) part of the irreducible bare susceptibility $\hat{\chi}_0$ at $\sigma_1 = \sigma_2 = \sigma_3 = \sigma_4$ ($\sigma_1 = \sigma_2 \neq \sigma_3 = \sigma_4$). $E_{\mathbf{k}}^n$ and $f(E_{\mathbf{k}}^n)$ are the eigenvalue and Fermi distribution function of Bogoliubov quasi-particles. $\hat{S}$ is the interaction vertex matrix[50]. We consider the on-site intra- and inter-orbital Coulomb interactions $U_l$ and $U'$ as $U_{d_{zx}/d_{yz}} = 0.21W$, $U_{d_{xy}} = 0.7 U_{d_{zx}/d_{yz}}$, $U' = 3 U_{d_{zx}/d_{yz}}/4$. The Hund's coupling and pair hopping are $J = J' = U_{d_{zx}/d_{yz}}/8$. We take the temperature $T = 1.9 \times 10^{-3} W$, the smearing factor $\delta = 1.9 \times 10^{-3} W$, and $1024 \times 1024 \times 32$ k-mesh. The resulting characteristic spectral features of the dynamical spin susceptibly are not much influenced by the detailed parameter values mentioned above.

**Acknowledgements**

The experiments at AMATERAS were conducted under the user program with the proposal numbers 2018A0060 and 2018AU1402. The present work was supported by JSPS KAKENHI Grant Numbers JP17K05553 and JP17K14349, and the Cooperative Research Program of "Network Joint Research Center for Materials and Devices" (20181072).

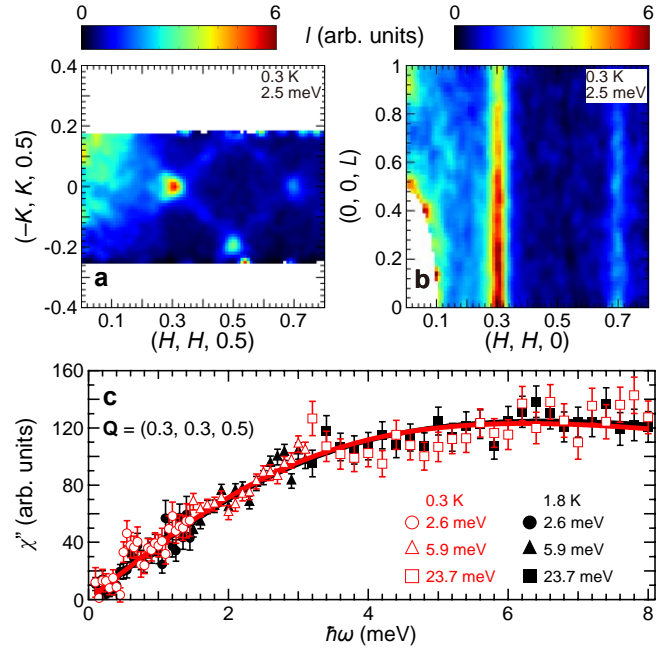

**Fig. 1** High-energy INS results of the IC magnetic fluctuations in Sr$_2$RuO$_4$. **a, b** Constant-energy INS intensity maps in the (**a**) (*HK*0.5) and (**b**) (*HHL*) planes with the energy window of [1.5, 3.5] meV at 0.3 K. **c** $\chi''(\hbar\omega)$ spectra at $\mathbf{Q}_{IC}$ = (0.3, 0.3, 0.5) below and above $T_c$. Data from different incident energies of neutrons $E_i$ = 2.64, 5.93, and 23.7 meV (circles, triangles, and squares) are combined for each temperature after background estimated by $\chi''(\hbar\omega)$ at **Q** = (0.525, 0.525, 0.5) is subtracted from each spectrum. Solid lines are fitting results by the conventional relaxation response model.

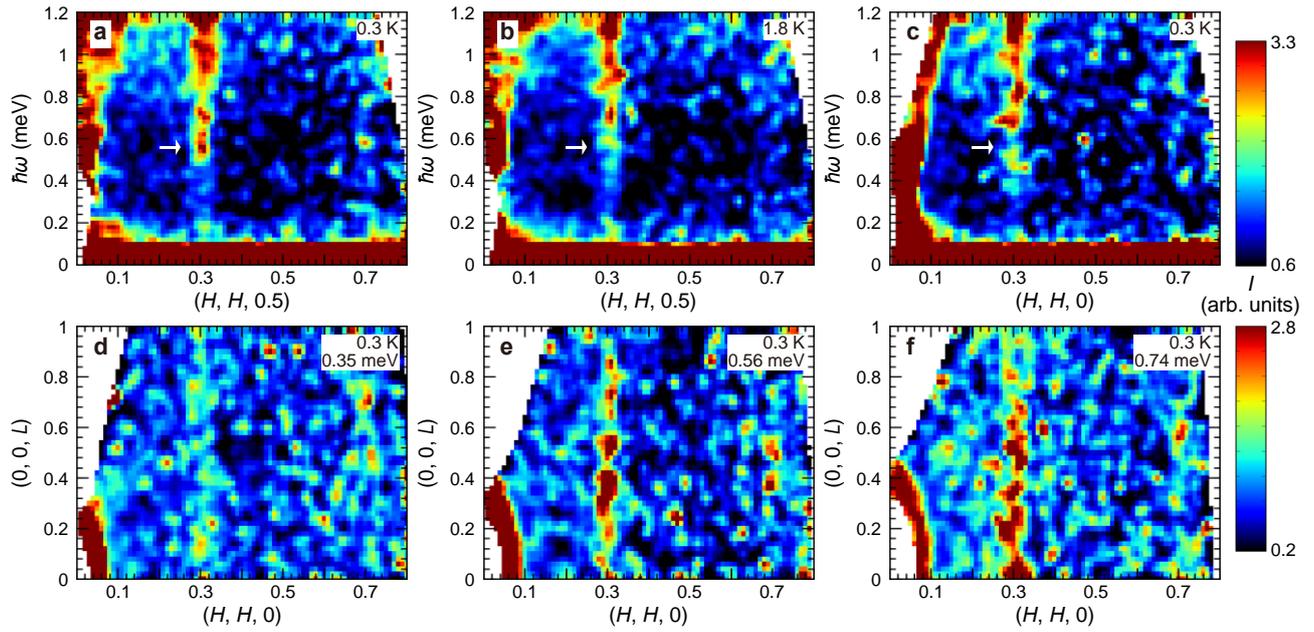

**Fig. 2** Low-energy INS intensity maps in Sr$_2$RuO$_4$. **a-c** Low-energy IC magnetic fluctuations (**a**) along **Q** = (*H*, *H*, 0.5) at 0.3 K, (**b**) along (*H*, *H*, 0.5) at 1.8 K, and (**c**) along (*H*, *H*, 0) at 0.3 K. White arrows in (**a-c**) indicate the energy of the superconducting gap 2Δ at **Q**$_{IC}$. **d-f** Constant-energy INS intensity maps in the (*HHL*) plane at 0.3 K with the energy windows of (**d**) [0.23, 0.47] meV, (**e**) [0.47, 0.65] meV, and (**f**) [0.65, 0.83] meV.

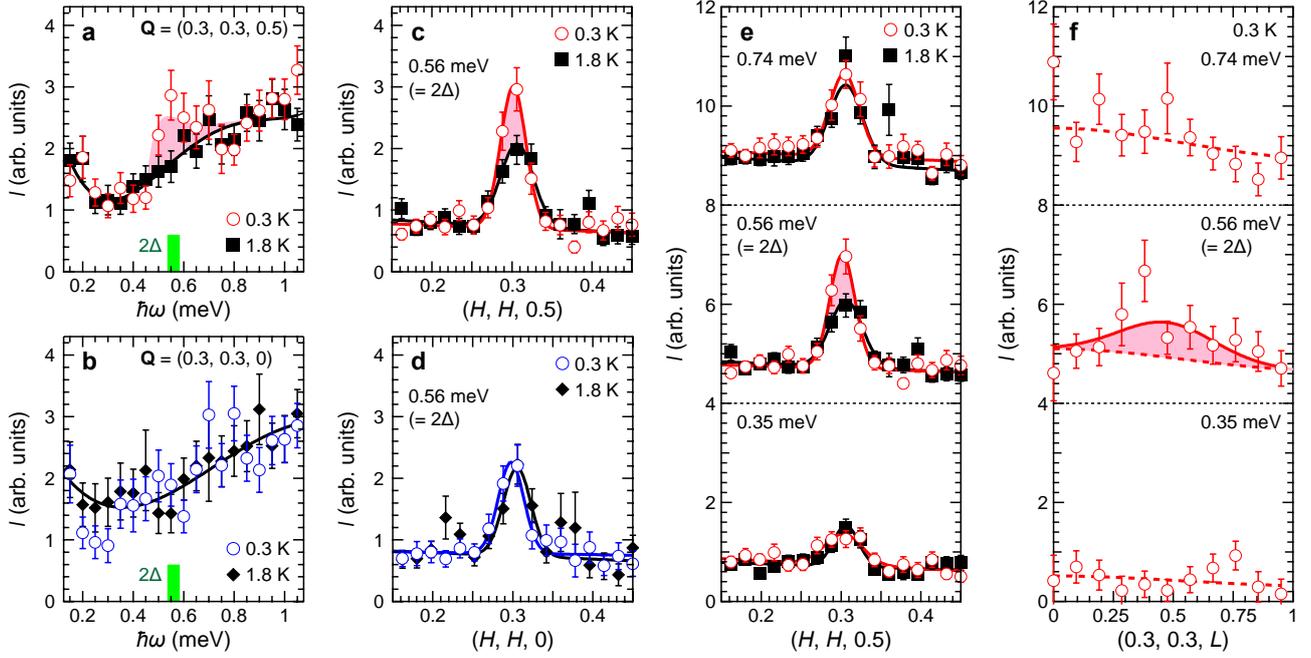

**Fig. 3** $I(\hbar\omega)$ and $I(\mathbf{Q})$ cuts of the low-energy IC magnetic fluctuations in $Sr_2RuO_4$. **a, b** $I(\hbar\omega)$ cuts at (**a**) $\mathbf{Q}$ = (0.3, 0.3, 0.5) and (**b**) (0.3, 0.3, 0) below and above $T_c$. Solid lines are the guides for the eye. Vertical vars represent the superconducting gap $2\Delta$ = 0.56 meV[37]. **c, d** $I(\mathbf{Q})$ cuts along (**c**) $\mathbf{Q}$ = (H, H, 0.5) and (**c**) (H, H, 0) at 0.3 and 1.8 K with the energy window of [0.47, 0.65] meV. Solid lines are the fitting results by the Gaussian function with linear background. **e** $I(\mathbf{Q})$ cuts along $\mathbf{Q}$ = (H, H, 0.5) at 0.3 and 1.8 K with the energy windows of [0.23, 0.47], [0.47, 0.65], and [0.65, 0.83] meV. Solid lines are the fitting results. **f** $I(\mathbf{Q})$ cuts along $\mathbf{Q}$ = (0.3, 0.3, L) at 0.3 K with the energy windows of [0.23, 0.47], [0.47, 0.65], and [0.65, 0.83] meV. Background estimated by $I(\mathbf{Q})$ along $\mathbf{Q}$ = (0.525, 0.525, L) at 0.3 K is subtracted from each spectrum. Dashed lines represent the squared magnetic form factor of $Sr_2RuO_4$[39] and the solid line is the guides for the eye. Solid areas in panels (**a,c,e,f**) indicate the spin resonance.

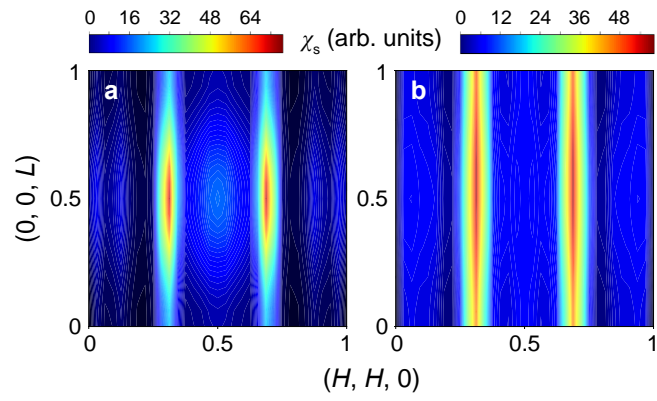

**Fig. 4** Calculated density maps of the dynamical spin susceptibility in $Sr_2RuO_4$. **a** $\omega = 2\Delta_0$. **b** $\omega = 4\Delta_0$.